\documentclass{article}
\usepackage[a4paper, total={175 mm, 260 mm}]{geometry}
\usepackage{blindtext}
\usepackage{multicol}
\usepackage{sectsty}
\sectionfont{\fontsize{12}{15}\selectfont}
\usepackage[utf8]{inputenc}
\usepackage{setspace}
\usepackage{lipsum}

\usepackage[english]{babel}
\usepackage{textcomp}
\usepackage{amsmath,amsfonts,amssymb}
\usepackage{graphicx}
\usepackage[shortlabels]{enumitem}
\usepackage{units} 
\usepackage{color} 
\PassOptionsToPackage{hyphens}{url}\usepackage{hyperref}
\usepackage{gensymb}
\usepackage{caption}
\usepackage{footnote}
\usepackage{float}
\usepackage[dvipsnames]{xcolor}
\usepackage{pdfpages}
\usepackage{subfiles}
\usepackage{subcaption} 
\usepackage{wrapfig}
\usepackage{pgfplots}
\pgfplotsset{compat=1.4}
\usepackage[toc,page]{appendix}

\usepackage{xcolor,colortbl}

\hypersetup{
    colorlinks=true, 
    linktoc=all,     
    linkcolor=blue,  
    citecolor = blue, 
    urlcolor=blue,
}

\usepackage{physics}

\usepackage[capitalise]{cleveref}

\usepackage{listings}
\definecolor{codegreen}{rgb}{0,0.6,0}
\definecolor{codegray}{rgb}{0.5,0.5,0.5}
\definecolor{codepurple}{rgb}{0.58,0,0.82}
\definecolor{backcolour}{rgb}{0.95,0.95,0.92}
 
\lstdefinestyle{mystyle}{
    backgroundcolor=\color{backcolour},   
    commentstyle=\color{codegreen},
    keywordstyle=\color{magenta},
    numberstyle=\tiny\color{codegray},
    stringstyle=\color{codepurple},
    basicstyle=\ttfamily\footnotesize,
    breakatwhitespace=false,         
    breaklines=true,                 
    captionpos=b,                    
    keepspaces=true,                 
    numbers=left,                    
    numbersep=5pt,                  
    showspaces=false,                
    showstringspaces=false,
    showtabs=false,                  
    tabsize=2
}
\usepackage{array}

\usepackage{nomencl}
\makenomenclature

\usepackage{etoolbox}
\renewcommand\nomgroup[1]{%
  \item[\Large\bfseries
  \ifstrequal{#1}{N}{Nomenclature}{%
  \ifstrequal{#1}{A}{Abbreviations}{}}%
]\vspace{10pt}} 

{}

\usepackage{titlesec}
\titleformat{\paragraph}
{\normalfont\normalsize\bfseries}{\theparagraph}{1em}{}
\titlespacing*{\paragraph}
{0pt}{3.25ex plus 1ex minus .2ex}{1.5ex plus .2ex}

\titlespacing\section{0pt}{12pt plus 4pt minus 2pt}{0pt plus 2pt minus 2pt}
\titlespacing\subsection{0pt}{12pt plus 4pt minus 2pt}{0pt plus 2pt minus 2pt}
\titlespacing\subsubsection{0pt}{12pt plus 4pt minus 2pt}{0pt plus 2pt minus 2pt}

\titlespacing*{\section}
{0pt}{0pt}{0pt}
\titlespacing*{\subsection}
{0pt}{0pt}{0pt}

\usepackage{tikz}

\usepackage{orcidlink}
\usepackage{dblfloatfix}
\usepackage{makecell}

\usepackage[
  style = chem-rsc,
  articletitle=true,
  sorting=none,
  backend=biber,
  autocite=superscript,
  giveninits=true,
  doi=true,
]{biblatex}
\DeclareNameAlias{author}{first-last}

\DeclareFieldFormat[article]{title}{#1}

\usepackage{booktabs}

\begin{document}
\begin{refsection}

\twocolumn
[\begin{center}
\Large{\textbf{Unravelling Challenges in Heating Power Measurements for Magnetic Hyperthermia - the RADIOMAG Round Robin Study Revisited}}\\
\normalsize
\vspace{3mm}
\textbf{\today}\\
\vspace{3mm}
Lise G. Hanson$^a$\orcidlink{0000-0002-8710-2610}, 
Daniel Ortega$^{b,c}$ \orcidlink{0000-0002-7441-8640},
Cathrine Frandsen$^{a,*}$\orcidlink{0000-0001-5006-924X}\\
\vspace{3mm}
$^a$Department of Physics, Technical University of Denmark, 2800 Kgs. Lyngby, Denmark\\
$^b$Condensed Matter Physics Department, University of Cádiz, 11510 Puerto Real, Spain\\ 
$^c$Biomedical Research and Innovation Institute of Cádiz, University of Cádiz, 11009 Cádiz, Spain\\
$^*$Corresponding authors, e-mail: \href{fraca@fysik.dtu.dk}{fraca@fysik.dtu.dk}
\end{center}
\vspace{1mm}

Non-adiabatic AC calorimetry is the most widely used technique for estimating the heating power of magnetic nanoparticles in magnetic hyperthermia. However, it is prone to systematic errors which lead to a standard deviation in the intrinsic loss power (ILP) of approximately 30–40\%, as revealed by the RADIOMAG EU COST Action TD1402 round-robin study involving 21 European laboratories. 
In this study, we re-examine the RADIOMAG dataset to both uncover previously unreported instrumentation issues, and to explore more deeply some of the reported instrumentation issues. We identify four common sources of error:
i) Insufficient temperature resolution, ii) AC-field sensitive thermometers, iii) Non-physical temperature oscillations, and iv) Apparent non-linear heat loss.
Based on these findings, we propose criteria for sufficient measurement quality and apply them to re-estimate the ILP values. These results have a standard deviation of 18–30\%, demonstrating that addressing instrumentation and analysis issues can improve measurement reliability and decrease the inter-laboratory deviation by up to 38\%.
When re-estimating ILP, we used the initial slope method, arguing that the corrected slope method, which was previously used to investigate the RADIOMAG data, could introduce misleading interpretations of systematic ILP deviations due to sub-optimal measurement conditions and the unnoticed influence of non-linear heat losses. However, we emphasize that the corrected slope method is preferred, given a linear heat loss.
Based on our analysis, we introduce a diagnostic protocol by using slope curves - a simple yet effective plot type - for identifying and solving common instrumentation challenges proactively before the data acquisition phase.

\vspace{3mm}]

\section{Introduction}

\begin{figure*}[b!]
    \centering
    \includegraphics[width=\linewidth]{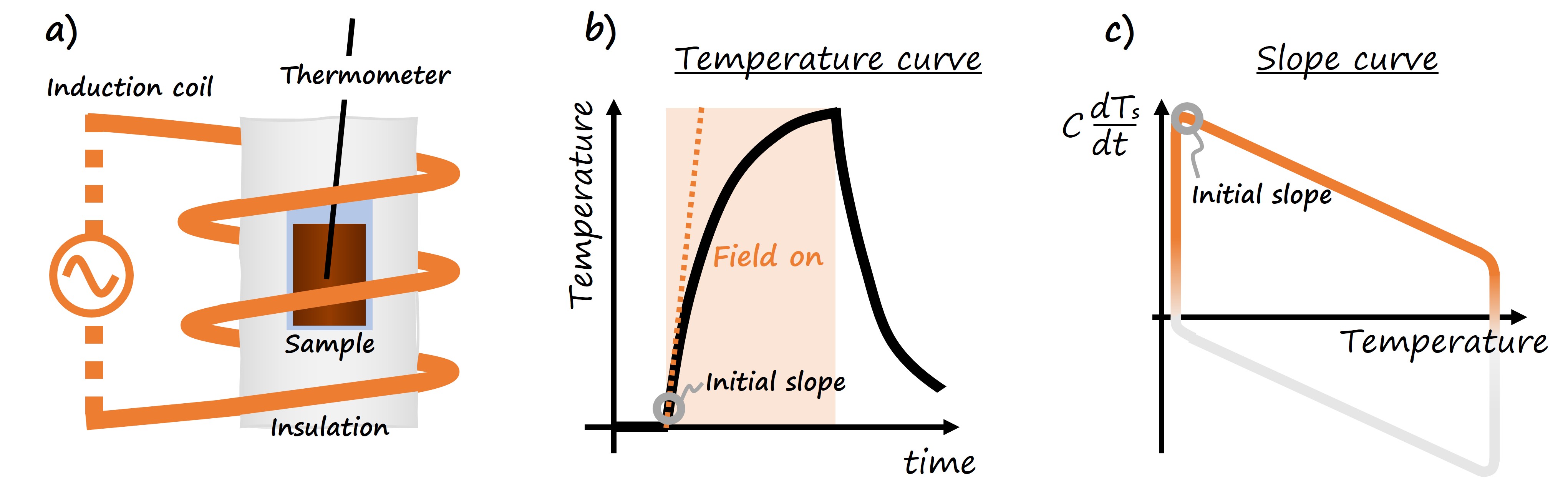}
    \caption{\textbf{a)} Illustration of a typical non-adiabatic calorimetric setup. \textbf{b)} Sample temperature before, during, and after field application in a non-adiabatic (solid line) and an adiabatic (dashed line) AC calorimetric setup. The orange area indicates when the field is applied. \textbf{c)} Slope curve for a linear heat loss. Orange indicates the heating phase when the field is on, i.e., $dT_s/dt>0$, and grey indicates the cooling phase with the field off, i.e., $dT_s/dt<0$.}
    \label{fig:Setup Tcurve SlopeCurve}
\end{figure*}

When magnetic nanoparticles are exposed to high-frequency magnetic fields, they can generate heat through induction heating due to magnetic hysteresis losses. This property enables a wide variety of applications, including cancer treatment (magnetic hyperthermia) \autocite{gilchrist1957selective,maier2011efficacy,chang2018biologically}, electrification of thermo-catalytic reactions \autocite{houlding2012application,mortensen2017direct,almind2020improving,adogwa2024catalytic,Bordet2025}, CO$_2$ capture and storage \autocite{gholami2022induction,sadiq2016magnetic,Schoukens2023}, and treatment of endometriosis \autocite{park2022targeted,moses2021nanomedicines,Talukdar2024}. 

To develop magnetic nanoparticles for such applications, it is essential to quantify their heating power. This allows for the characterization and comparison of nanoparticle samples and the calculation of application-specific parameters, such as the required medical dose.
However, in 2021, results from the EU COST action RADIOMAG showed systematic inter-laboratory deviations of $\pm$30-40\% in the intrinsic loss power (field and nanoparticle mass normalised heating power) across 17 state-of-the-art European laboratories using
non-adiabatic AC calorimetry \autocite{wells2021challenges}.  This is a significant issue, as non-adiabatic AC calorimetry remains the most accessible and widely used method for estimating heating power \autocite{andreu2013accuracy}. 

Several efforts have been undertaken to reduce deviations in measured heating powers, including considerations regarding vial material, sample volume, field homogeneity, thermometer position, field application schemes, data processing methods, temperature dependence of the heating power, and insulation effects \autocite{wildeboer2014reliable,huang2012measurement,wells2021challenges,papadopoulos2020magnetic,landi2013simple,Hanson2023impact,carlton2023new, Ruta2024}. Despite these efforts, discrepancies persist.

In the RADIOMAG project, identical magnetic nanoparticle samples were distributed to 21 laboratories for a round-robin test. Each laboratory performed AC calorimetric measurements on the samples and submitted their results for centralised analysis \autocite{wells2021challenges}. The RADIOMAG project thus gives a unique snapshot of the current state of the art of AC calorimetry. 

Due to variations in instrumentation, the applied field amplitude, $H_0$, and frequency, $f$, differed across laboratories. To enable comparison of measurements across the different laboratories, a field-independent normalised heating power known as the intrinsic loss power (ILP) was used
\begin{gather}
    \text{ILP} = \frac{P_\text{MNP}}{m_\text{MNP} H_0^2 f},
    \label{eq:ILP}
\end{gather}
where $P_\text{MNP}$ is the heating power of the magnetic nanoparticles, and $m_\text{MNP}$ is the mass of the magnetic nanoparticles in the sample.
ILP is expected to be field independent if linear-response theory applies \autocite{carrey2011simple}. The RADIOMAG samples were synthesised to meet the criteria of linear-response theory \autocite{wells2021challenges}. 


In this article, we re-analyze the RADIOMAG dataset \autocite{zenodoWells} to identify instrumentation-related issues in the current state of AC calorimetry. We categorize four primary sources of error: 
\begin{enumerate}[i)]
    \item Low temperature resolution 
    \item AC-field sensitive thermometers
    \item Non-physical heat oscillations
    \item Apparent non-linear heat losses
\end{enumerate}
We assess the impact of each issue on ILP estimation and propose a set of criteria for evaluating data quality. Hereafter, ILP values are re-estimated for measurements that meet these criteria.
The revised ILP values show deviations of $\pm$18–30\%, representing an improvement up to 38\% when comparing with the deviations reported by Wells et al.\autocite{wells2021challenges}. Finally, we propose a diagnostic protocol for identifying instrumentation issues in non-adiabatic AC calorimetry setups. 


\section{Estimating heating power}

In this section, we review how adiabatic and non-adiabatic AC calorimetry is used to estimate $P_\text{MNP}$, which is required to determine the ILP value of the magnetic nanoparticle samples, see \cref{eq:ILP}. 

AC calorimetry is based on measuring the temperature evolution in a sample while an alternating magnetic field is applied. The applied field causes the nanoparticles to heat up, which increases the sample temperature. For magnetic hyperthermia applications, the sample consists of magnetic nanoparticles dispersed in liquid or cell medium. 
In the adiabatic case, the sample is attempted to be perfectly insulated, and no heat losses occur. In contrast, in non-adiabatic setups, heat losses are present. \cref{fig:Setup Tcurve SlopeCurve}a illustrates a typical non-adiabatic AC calorimetry setup.

In both adiabatic and non-adiabatic AC calorimetry, the master equation for estimating $P_\text{MNP}$ is given by the lumped heat capacity model
\begin{gather}
    C\frac{dT_s}{dt}=P_\text{MNP}-P_\text{loss},
    \label{eq:LumpedHeatCap}
\end{gather}
where $C$ is the heat capacity of the sample, $T_s$ is the sample temperature, $t$ is time, and $P_\text{loss}$ is the heat loss from the sample to the surroundings. 

In the adiabatic case, $P_\text{loss}=0$, and thus the sample temperature increases linearly, assuming $P_\text{MNP}$ is constant (\cref{fig:Setup Tcurve SlopeCurve}b dashed line). Obtaining an adiabatic setup is instrumentally challenging and thus only a few adiabatic AC calorimetric setups have been reported \autocite{andreu2013accuracy,wells2021challenges}. 

In the non-adiabatic AC calorimetry, it is necessary to account for the term $P_\text{loss}$. Two common methods are: i) The initial slope method, where the heat loss is neglected, and ii) The corrected slope method, where a linear heat loss is assumed. 

\subsection{Initial slope method}
At the onset of the heating process, the temperature difference between the sample and the surroundings is very small, and the heat loss, $P_\text{loss}$, is therefore expected to be negligible. In this case, the heating power can be determined by
\begin{gather}
    P_\text{MNP} = C\frac{d T_s}{dt}\Big|_{t\approx 0}\,.
    \label{eq:initialSlope}
\end{gather}
Typical heating curves for non-adiabatic and adiabatic measurements are shown in \cref{fig:Setup Tcurve SlopeCurve}b. Initially, the curves overlap due to the negligible heat loss. The initial slope method is known to underestimate the heating power of the magnetic nanoparticles \autocite{wildeboer2014reliable,CABRERA2019111}, see also \cref{sec:Slope curves}.

\subsection{Corrected slope method}
The corrected slope method was developed by Wildeboer et al. \autocite{wildeboer2014reliable} and was used by Wells et al. to analyse the RADIOMAG data \autocite{wells2021challenges}.
\begin{figure}[b!]
    \centering
    \includegraphics[width=0.8\linewidth]{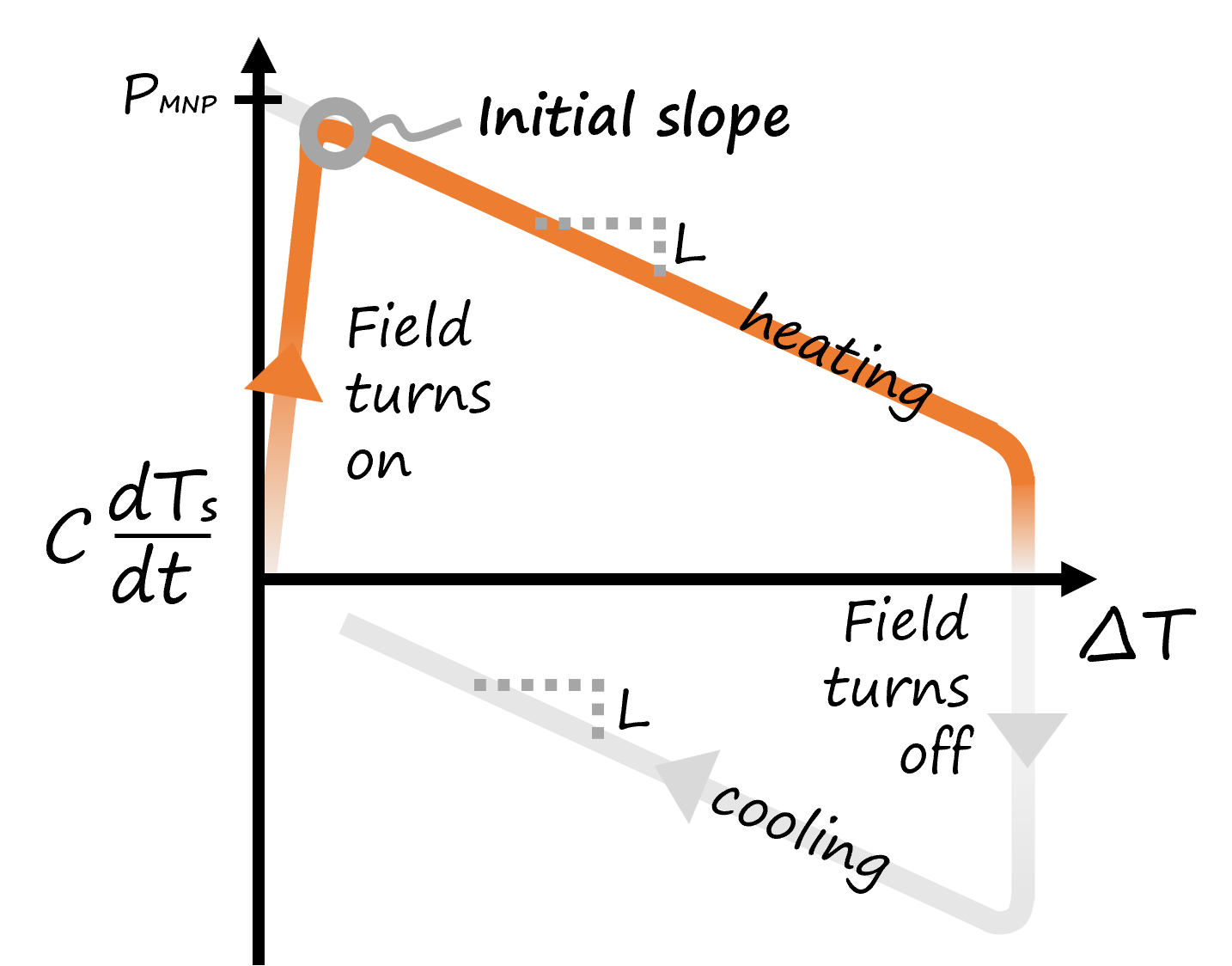}
    \caption{Slope curve for a linear heat loss. Orange indicates the heating phase, and grey indicates the cooling phase. Both the heating and cooling phases have a slope of $L$. $P_\text{MNP}$ indicates the true nanoparticle heating power. The arrows indicate the direction of time.}
    \label{fig:slopeCurve}
\end{figure}

The method is based on the assumption that $P_\text{loss}$ increases linearly with the temperature difference between the sample and surroundings, $\Delta T$, often referred to as Newton's law of cooling. In this case, \cref{eq:LumpedHeatCap} becomes
\begin{gather}
     P_\text{MNP}=C\frac{dT_s}{dt} + L\Delta T,
     \label{eq:CorrectedSlope}
\end{gather}
where $L$ is a constant called the linear loss parameter. In most studies, the temperature of the surroundings is taken as the thermal equilibrium temperature of the sample before the field is applied. However, this has been shown to be a bad assumption, especially for insulated samples \autocite{Hanson2023impact}. The problem is that the temperature of the sample insulation increases with the heating of the sample, and therefore $\Delta T$ is erroneously defined, and a linear heat loss can appear non-linear \autocite{Hanson2023impact}. This can be resolved by measuring the actual $\Delta T$ or by avoiding a temperature increase in the surroundings, e.g., by air flow and no insulation \autocite{Hanson2023impact}. If the actual $\Delta T$ is included in the data analysis, then the corrected slope method is adequate \autocite{Hanson2023impact} and preferred over the initial slope method as it should be more accurate \autocite{wildeboer2014reliable}.

\subsection{Slope curves}
\label{sec:Slope curves}
The analysis presented in this article is founded on slope curves, defined as plots of $C\cdot dT_s/dt$ vs. $\Delta T$ (or $T_s$) as illustrated in \cref{fig:Setup Tcurve SlopeCurve}c. The strength of the slope curves lies in their visualisation of $dT_s/dt$, which is the fundamental parameter in both the initial and corrected slope methods, see \cref{eq:LumpedHeatCap,eq:CorrectedSlope,eq:initialSlope}.
Thus, these curves are an ideal tool for inspecting AC calorimetric data, including the linearity of heat losses. 

When the AC magnetic field is applied, the temperature increases ($C\cdot dT_s/dt>0$), as illustrated with the orange curve in \cref{fig:slopeCurve}. When the field is removed, the sample cools back to the baseline temperature ($C\cdot dT_s/dt<0$) as shown by the gray curve. A more thorough explanation of slope curves and their interpretation is given in Hanson et al. \autocite{Hanson2023impact}.

From the slope curve perspective, the initial slope method corresponds to estimating $P_\text{MNP}$ from the value of $C\cdot dT_s/dt$ just after the field is turned on, as indicated with the gray circle in \cref{fig:slopeCurve}. The initial slope method underestimates the heating power because the experimental slope cannot be evaluated exactly at $\Delta T = 0$, as multiple data points are required to numerically calculate the slope. This underestimation is illustrated by the extended gray line near the initial slope point, which extrapolates $P_\text{MNP}$ back to the value at $\Delta T =0$.

The corrected slope method can also be visualised using slope curves. Plotting \cref{eq:CorrectedSlope} as $C\cdot dT_s/dt = P_\text{MNP}-L\Delta T$ yields a linear slope curve, as shown in \cref{fig:slopeCurve}. The linear loss parameter $L$ is obtained from a linear fit to the heating or cooling data in the slope curve (preferably the cooling data). Adjusting for heat loss allows extrapolation of the slope curve back to $\Delta T = 0$, where the intercept corresponds to $P_\text{MNP}$.
If the heat loss is non-linear with respect to $\Delta T$, or if $\Delta T$ is erroneously defined, then the slope curves will deviate from linearity \autocite{Hanson2023impact}.  



\section{Methods}
\label{sec:Methods}
This section outlines the methods and samples from the RADIOMAG project relevant to this study. A more extensive description of the measurement protocols, sample selection, and study design can be found in Wells et al. \autocite{wells2021challenges}. 

In total, three rounds of AC calorimetric measurements were performed in the RADIOMAG project \autocite{wells2021challenges}. Each round consisted of sample delivery, calorimetric measurements, and submission of measured data and instrument specifications. In each round, measurements were repeated three times per sample.

\textbf{Round 1:} For the 1st round, two batches of different samples were synthesised, named Sample 1 and Sample 2. Portions of 2 ml of each sample were distributed to the participating laboratories. Both samples consisted of ferrimagnetic iron-oxide nanoparticles in aqueous suspensions, with concentrations shown in \cref{tab:concentrations}. Data were obtained from 21 laboratories. 

\begin{table}[b]
    \centering
    \begin{tabular}{c c c}
    \hline
    \rowcolor{lightgray!60}
    Sample name   & $c_\text{MNP}$ [mg MNP/ml] & Material$^*$  \\ \hline
    Sample 1     & 40.6 & $\gamma$-Fe$_2$O$_3$\\
    Sample 2    & 3.9  & Fe$_3$O$_4$ \\ \hline
    \end{tabular}
    \caption{Overview of the magnetic nanoparticle concentration, $c_\text{MNP}$, and the primary$^*$ material of the nanoparticles.}
    \label{tab:concentrations}
\end{table}

\textbf{Round 2:} In the 2nd round, new 2 ml portions of Samples 1 and 2 were delivered and measured in the participating laboratories. Additionally, a sample of zero conductivity water was distributed to each laboratory (nominal 0 mS/cm). The measurement protocol was refined regarding thermometer positioning, and an upper temperature limit of 60$\degree$C was introduced. Data were obtained from 12 laboratories. 

\textbf{Round 3:} In the 3rd round, new sample batches were used, and a sample of zero conductivity water was again distributed to all laboratories. In this study, only the water data from round 3 are used.
Data were obtained from 15 laboratories. 

\begin{figure*}[t]
    \centering
    \includegraphics[width=\linewidth]{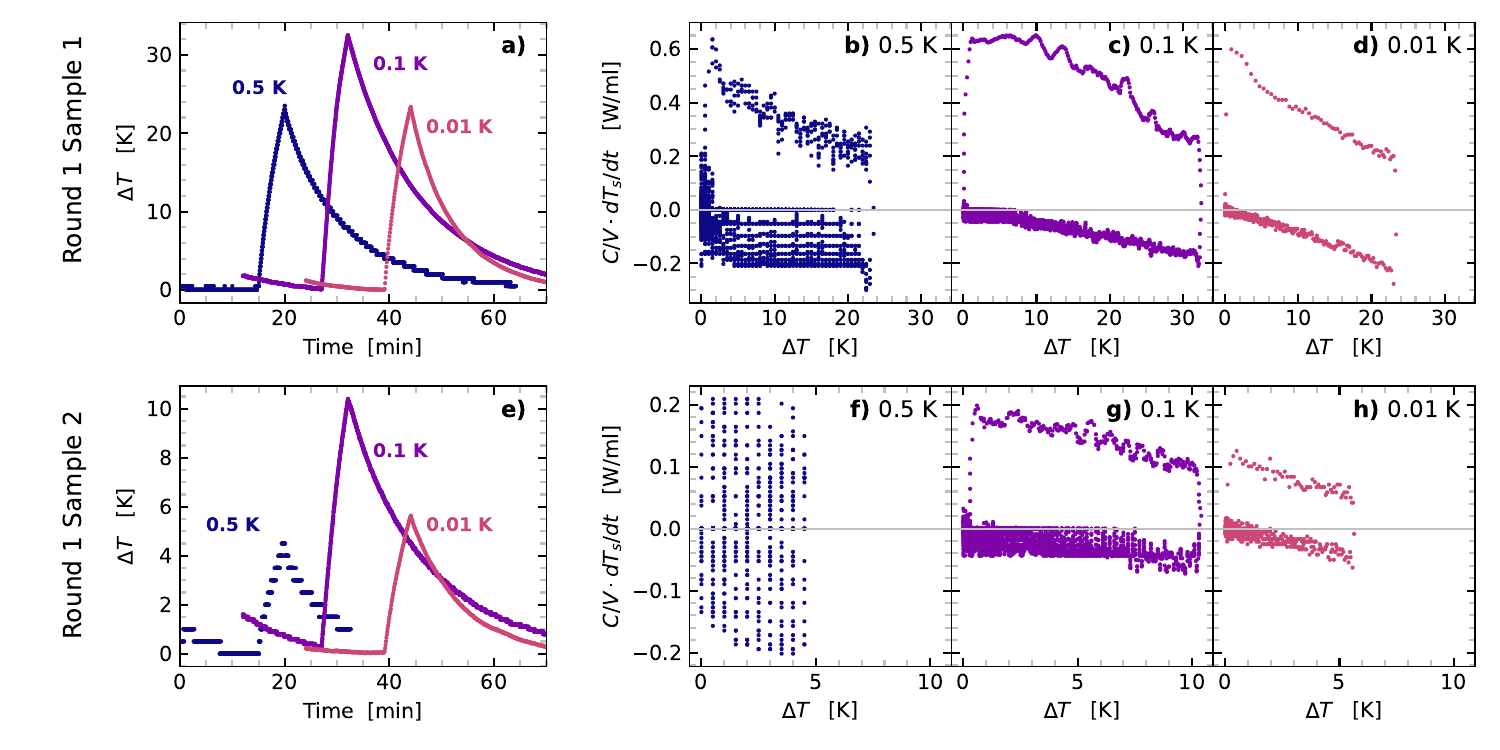}
    \caption{Examples of data with different temperature resolution, $\delta T_s$, from round 1 measured on samples 1 (top panels) and 2 (bottom panels).\textbf{ a} and \textbf{e)} Show sample temperatures as a function of time. \textbf{b, c, d, f, g,} and \textbf{h)} Show the corresponding slope curves. The colours refer to the laboratory numbers: Lab 2 (blue), 14 (purple), and 21 (pink). }
    \label{fig:TresPlot}
\end{figure*}

The applied fields varied across laboratories, with amplitudes ranging from 3 to 15.7 kA/m and frequencies from 111.7 to 928 kHz. All laboratories used a non-adiabatic setup, except laboratory 19, which used an adiabatic setup. The measurement protocol requested a sampling rate of 1 s, but many laboratories used different rates. In our analysis, we accepted these variations as inherent to the setup differences and did not interpolate the data to a uniform 1 s rate. The laboratory numbers used in this article follow the same numbering as used in Wells et al. \autocite{wells2021challenges} except for laboratory 18-21 which were excluded from their analysis. The heat capacity of all samples is assumed to be that of water, justified by the low nanoparticle concentration.


A Savitzky-Golay filter was used to determine the time derivatives and to reduce the noise originating from numerical derivatives. The filter was applied across the entire dataset, so interpolation only occurs at the very beginning and end of each measurement. 
In all of the data analysis, the filter has been fixed to a window width of $W=15$ s, and a polynomial order of 1. The typical field application times were around 5 min.

For the initial slope method, we used the maximum slope
\begin{gather}
    P_\text{MNP} = C\max\left(\frac{d T_s}{dt}\right).
\end{gather}
The maximum is chosen as it is expected that values lower than this are lower due to either the influence of heat loss ($t\gtrsim W$) or initialisation ($t\lesssim W$). Noise may cause these points to become artificially high, but the Savitzky-Golay filter width of $W=15$ s removes significant effects of noise. 

In all slope curves presented below, the unit on the y-axis is given in W/ml, since the sample volume, $V$, varies between laboratories. This is determined by 
\begin{gather}
    \frac{C}{V}\,\frac{dT_s}{dt} \,.
\end{gather}

ILP values from the adiabatic setup in laboratory 19 were determined using a method corresponding to the corrected slope method, see \cref{SI:ILP adiabatic}.

\section{Results and discussion}
\subsection{Measurement issues}
Four classes of instrumentation issues are identified and discussed in this section. Slope curves are essential for analysing all datasets, and slope curves from all laboratories are shown in \cref{SI:Slope curves}.

\subsubsection{Temperature resolution}\label{sec:Tres}
Many laboratories did not give extensive information about the model of their thermometers which makes it impossible to know the true temperature resolutions of all laboratories. Instead, we will consider the temperature resolution as the smallest non-zero temperature step in the data, defined as
\begin{gather}
    \delta T_s =\text{min}(|T_{s,i+1}-T_{s,i}|) \, ,
\end{gather}
where $i$ indexes the data points in a measurement. 

\cref{tab:Tres} provides an overview of the distribution of $\delta T_s$ within all 21 laboratories. It is evident that $\delta T_s$ varies with 2 orders of magnitude from $\leq$ 1 mK to 0.5 K. 
\begin{table}[t]
    \centering
    \begin{tabular}{ccccc}
    \hline
    \rowcolor{lightgray!20}
        $\delta T_s$ [K]  & 0.5 &  0.1 & 0.01 & $\leq$0.001\\
        \hline
        Number of Laboratories & 2 & 9 & 8 &  2\\
        \hline
    \end{tabular}
    \caption{Overview of the distribution of $\delta T_s$ values in all 21 laboratories participating in the RADIOMAG project.
    }
    \label{tab:Tres}
\end{table}

\begin{figure}[b!]
    \centering
    \includegraphics[width=\linewidth]{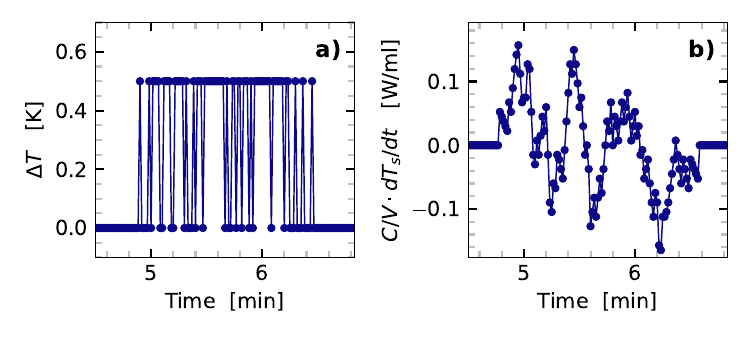}
    \caption{Data from Lab 2 ($\delta T_s=0.5$ K) Sample 1 before the field is applied. The data is the same as in \cref{fig:TresPlot}a. \textbf{a)} Raw data temperature data before the field is applied, \textbf{b)} Time derivative $C/V\cdot dT_s/dt$ calculated from the data in \cref{fig:T jumps}a.}
    \label{fig:T jumps}
\end{figure}

\cref{fig:TresPlot} show data from three laboratories with $\delta T_s$ values of 0.5, 0.1, and 0.01 K. \cref{fig:TresPlot}a show the directly measured calorimetric data on Sample 1. At first glance, $\delta T_s$ does not appear to significantly affect the smoothness of the temperature curves. However, from the corresponding slope curves (\cref{fig:TresPlot}b-d), it becomes clear that the numerical derivative is much noisier for $\delta T_s = 0.5$ K compared to lower $\delta T_s$-values.

\cref{fig:TresPlot}e shows calorimetric measurements from the same laboratories on Sample 2.  Comparing panels a) and e), the temperature increase for Sample 2 is about 3–4 times lower than for Sample 1, due to its lower nanoparticle concentration and difference in nanoparticle types. The data points with $\delta T_s=0.5$ K appear clearly discrete, caused by the relatively small temperature increase of approximately 5 K. \cref{fig:TresPlot}f-h show slope curves for Sample 2. The slope curve with $\delta T_s = 0.5$ K is too noisy to interpret, due to the combination of a high value of $\delta T_s$ and a slow heating. Thus, when evaluating whether the thermometer resolution is sufficient, it is not the absolute value of $\delta T_s$ that matters, but its value relative to the rate of temperature increase. 

\cref{fig:T jumps}a shows how the high value of $\delta T_s = 0.5$ K causes the temperature to fluctuate between two discrete values of $T_s$ before the field is turned on. \cref{fig:T jumps} b) shows that this leads to fluctuations in $C/V\cdot dT_s/dt$ on the order of 0.1 W/ml, which is a considerably high value compared to the maximal value of $C/V\cdot dT_s/dt$ (initial slope method) being 0.6 W/ml, as seen in \cref{fig:TresPlot}e. 

The high level of fluctuations in $dT_s/dt$ is problematic, since $dT_s/dt$ is essential in both the initial slope method and the corrected slope method. Moreover, the noise makes it difficult to assess the linearity of heat losses, which is necessary to validate the corrected slope method. Smoothing the noise by widening the Savitzky-Golay filtering window is undesirable since the filter would smooth out real features in the data. 


In conclusion, the temperature resolution must be sufficiently high in relation to the rate of increase in temperature. As a general laboratory practice, the temperature should not be rounded to the instrument uncertainty, but more digits should be kept. For example, we have experienced that older versions of the nanoTherics software (the brand of several AC calorimetric devices and used by five of the participating laboratories) round the temperature to the uncertainty, although the sensor (Onsensa Innovations, PRB-G40) records with more digits.

As a criterion for a sufficient temperature resolution, we introduce the following limit
\begin{gather}
    \left. \frac{\delta T_s}{W}\middle/\left[\frac{dT_s}{dt}\right]_\text{max} \right. <10\,\%,
    \label{eq:dTs criteria}
\end{gather}
where $W$ is the width of the Savitzky-Golay filter, here 15 s. The term $\delta T_s/W$ represents the maximal slope that could be caused by the fluctuations. The term $[dT_s/dt]_\text{max}$ is the maximal slope observed in the data. The limit of 10\% was chosen by qualitative investigation of the slope curves. The rejected laboratories based on Eq. \eqref{eq:dTs criteria} are
\begin{itemize}
    \item Sample 1: Laboratory 2
    \item Sample 2: Laboratory 1, 2, 14, 15 and 17
\end{itemize}

More measurements are rejected in the case of Sample 2 due to its lower heating rate. See \cref{SI:T res criteria} for an overview of the calculated values of \cref{eq:dTs criteria} across all laboratories.



\subsubsection{Thermometers and field influence}
\begin{figure}[t!]
    \centering
    \includegraphics[width=\linewidth]{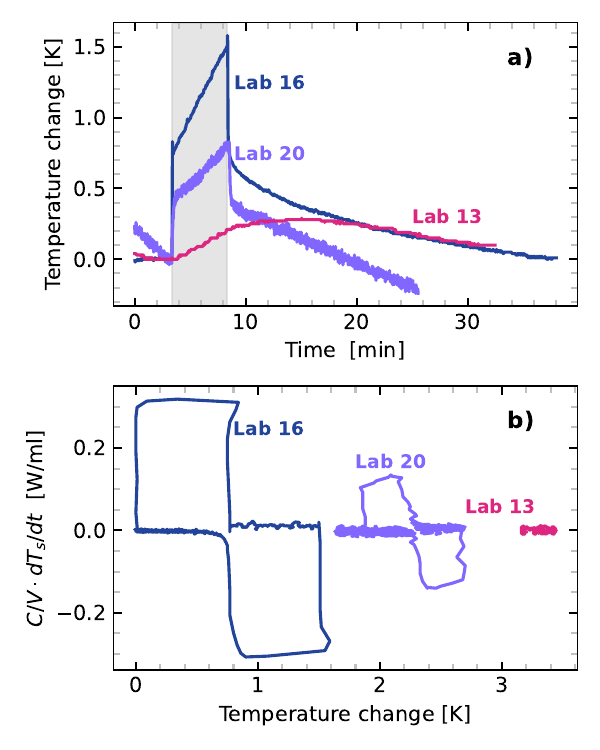}
    \caption{Examples of measurements in water for investigating AC field sensitivity of the AC calorimetric setups. \textbf{a)} Measured temperature as a function of time. The grey area indicates when the field is on. \textbf{b)} Corresponding slope curves. A shift in the temperature change is introduced to avoid the plots from overlapping.}
    \label{fig:Thermometer sensitivity to AC field}
\end{figure}

 Optical thermometers are known not to be field sensitive, whereas thermocouples must be used with caution when exposed to high-frequency AC fields. All thermocouples are made of metal, and thus heat is generated by induced eddy currents, scaling with the size of the thermometer. While metallic segments of tens of microns thick may remain stable, those with thicknesses of several hundred microns already show severe heating problems \autocite{Chan1993}.
Additionally, the thermocouple is susceptible to electromagnetic interference from the coil, which causes faulty readings if not handled correctly. Thus, in the case of thermocouples, it must be confirmed how the field application affects the thermometer. Out of the 21 laboratories, 16 used optical thermometers and 5 used thermocouples (labs 1, 2, 16, 20, and 21). 

A simple method to investigate the AC field sensitivity of a thermometer is to measure the temperature of a sample which does not produce heat under induction, e.g., water, during AC field application. The sample temperature is not expected to increase, except for potential temperature changes originating from external sources such as Joule heating in the induction coil. 
From rounds 2 and 3, AC calorimetric reference measurements on deionised water were obtained from all laboratories except laboratories 1 and 5. 

\begin{figure*}[t!]
    \centering
    \includegraphics[width=\linewidth]{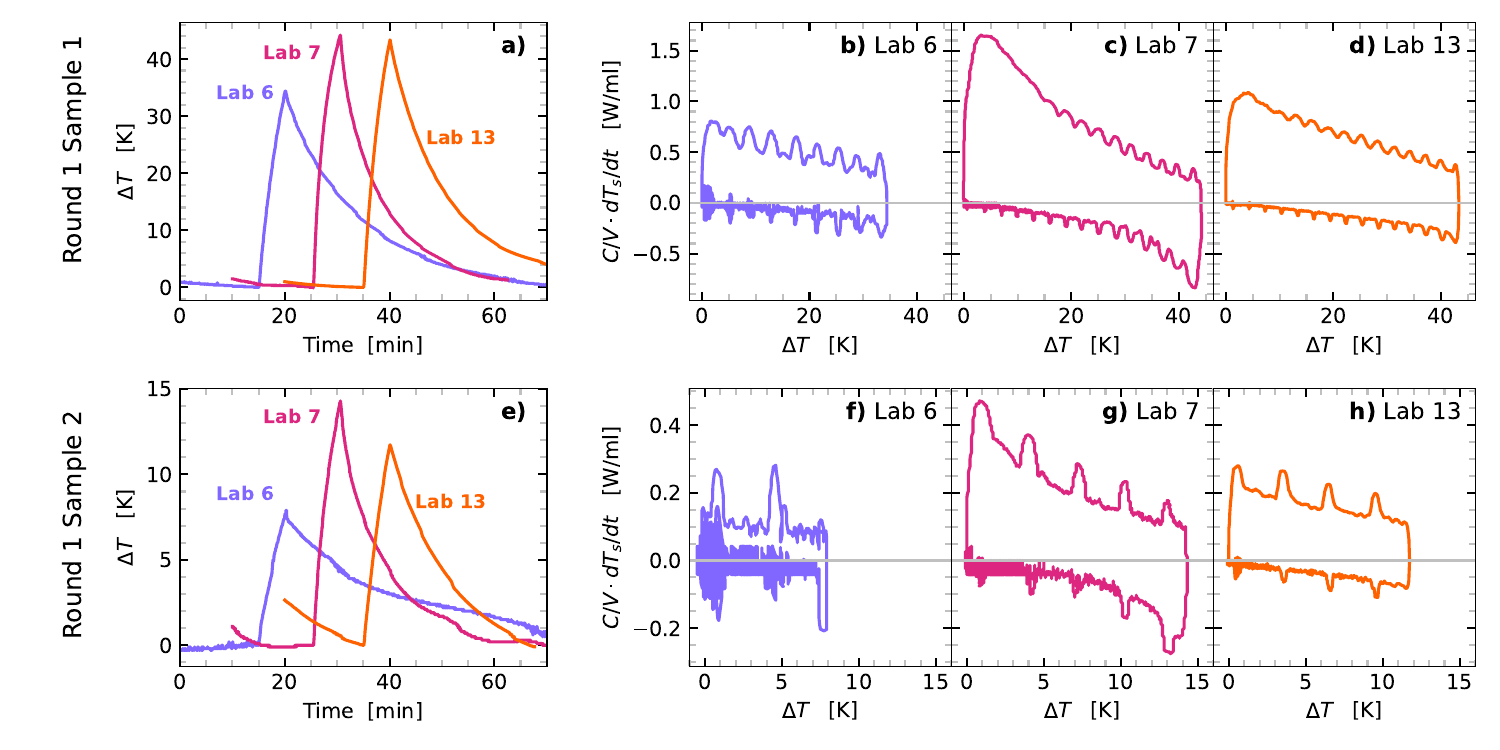}
    \caption{Examples of measurements where non-physical temperature oscillations occur. The data are from round 1 and measured on Samples 1 (top panels) and 2 (bottom panels). \textbf{a} and \textbf{e)} Show the measured temperature change as a function of time. \textbf{b, c, d, f, g,} and \textbf{h)} Show the corresponding slope curves.}
    \label{fig:T oscillations}
\end{figure*}

\cref{fig:Thermometer sensitivity to AC field}a shows AC calorimetric data measured on water from three laboratories. For laboratory 16 and 20 using thermocouples of type T, the temperature jumps rapidly when the field is turned on, followed by a slower increase. When the field is turned off, a sharp decrease in temperature is seen, and then the temperature decreases at a slower pace. It is concluded from \cref{fig:Thermometer sensitivity to AC field}a, that the thermometers used in laboratories 16 and 20 are affected by the AC field. The temperature jump is expected to be caused by electromagnetic interference, while the slow temperature increase is caused by eddy currents. For laboratory 13 using optical fibre measurements, only a slow temperature change is observed, probably caused by a temperature increase and subsequent decrease of the sample environment due to the induction coil. 

\cref{fig:Thermometer sensitivity to AC field}b shows the corresponding slope curves. 
The slope curve from laboratory 16 reaches the highest value of $C/V \cdot dT_s/dt$, indicating that its thermometer sensitivity has the greatest influence on the measured heating power among the three laboratories. For laboratory 13, no sharp temperature changes are observed, and thus the slope remains low, i.e., the measured heating power is insignificantly affected by a slow temperature change. 

No water measurements were provided from laboratory 1 and 5. Instead, slope curves from Sample 2 were investigated, as its heating power is low. Laboratory 1 was found to suffer from a field-sensitive thermometer based on non-physical bumps in the slope curve, similar to those seen for laboratory 16 and 20 in \cref{fig:Thermometer sensitivity to AC field}b, see \cref{SIfig:Slope curve lab 1}. 

In summary, laboratories 1, 16, and 20 were found to use field-sensitive thermometers. This corresponds with the fact that all three laboratories used thermocouples instead of optical fibres. Laboratories 2 and 21 also used thermocouples. The temperature resolution of laboratory 2 was insufficient, and thus, no conclusion could be reached on AC field sensitivity. In laboratory 21, the measurements were not significantly affected by the AC field, showing that thermocouples can be used in AC calorimetry, but their field sensitivity must be evaluated beforehand to ensure reliable temperature readings. 

As a criterion for sufficiently low thermometer AC-field influence and temperature increase caused by external effects (e.g. the coil), the height of the jump measured in water must not affect the slope $dT_s/dt$ with more than 10\%. 
\begin{gather}
    \left. \left[\frac{dT_s}{dt}\right]_\text{max}^\text{Water}\middle/\left[\frac{dT_s}{dt}\right]_\text{max}^\text{Sample} \right. <10\,\%.
    \label{eq:AC and water criteria}
\end{gather}
This limit is only exceeded for laboratories 1, 16 and 20  due to the AC-field influence.

\subsubsection{Non-physical heat oscillations}
In this section, we investigate an effect which we refer to as non-physical heat oscillations. To our knowledge, these oscillations have not previously been described in the field of AC calorimetry for magnetic nanoparticles.

\cref{fig:T oscillations} shows examples of data where non-physical heat oscillations occur. The oscillations consist of periodically stronger and weaker heating, as seen from the slope curves. We consider this oscillatory heating and cooling to be non-physical, as to our knowledge, no known mechanism in the system would cause heating and cooling to become stronger or weaker in temperature periodic bursts. The oscillations are problematic because they can cause over- or underestimation of $dT_s/dt$, thereby affecting the ILP value. 

Comparing \cref{fig:T oscillations}b-d vs. f-h, it is evident that the oscillations are relatively stronger in the slope curves for Sample 2 with lower heating power than for Sample 1, since the oscillation height is preserved. Thus, the oscillations influence the estimates on the heating power more for Sample 2 than for Sample 1. 

The laboratories that exhibited non-physical heat oscillations are laboratories 5, 6, 7, and 13. All of these laboratories used optical fibres for measuring the temperature. Specifically, two of these laboratories used a sensor from Neoptix (T1C-01-PP10 lab 5 and Reflex 4 lab 13), which was not used by anyone else. Thermometer names were not provided by laboratories 6 and 7. It is observed that the oscillations are evenly spaced in temperature, with approximately one oscillation occurring for every $\Delta T_s$ of 3$\degree$C. This periodicity is consistent across all four laboratories, showing non-physical heat oscillations. 

As a criterion for sufficiently small non-physical heat oscillations, the oscillation height (found by visual inspection of the slope curves) must be less than $10\%$ of $[dT_s/dt]_\text{max}$.

\subsubsection{Apparent non-linear heat losses}
\begin{figure*}[t!]
    \centering
    \includegraphics[width=\linewidth]{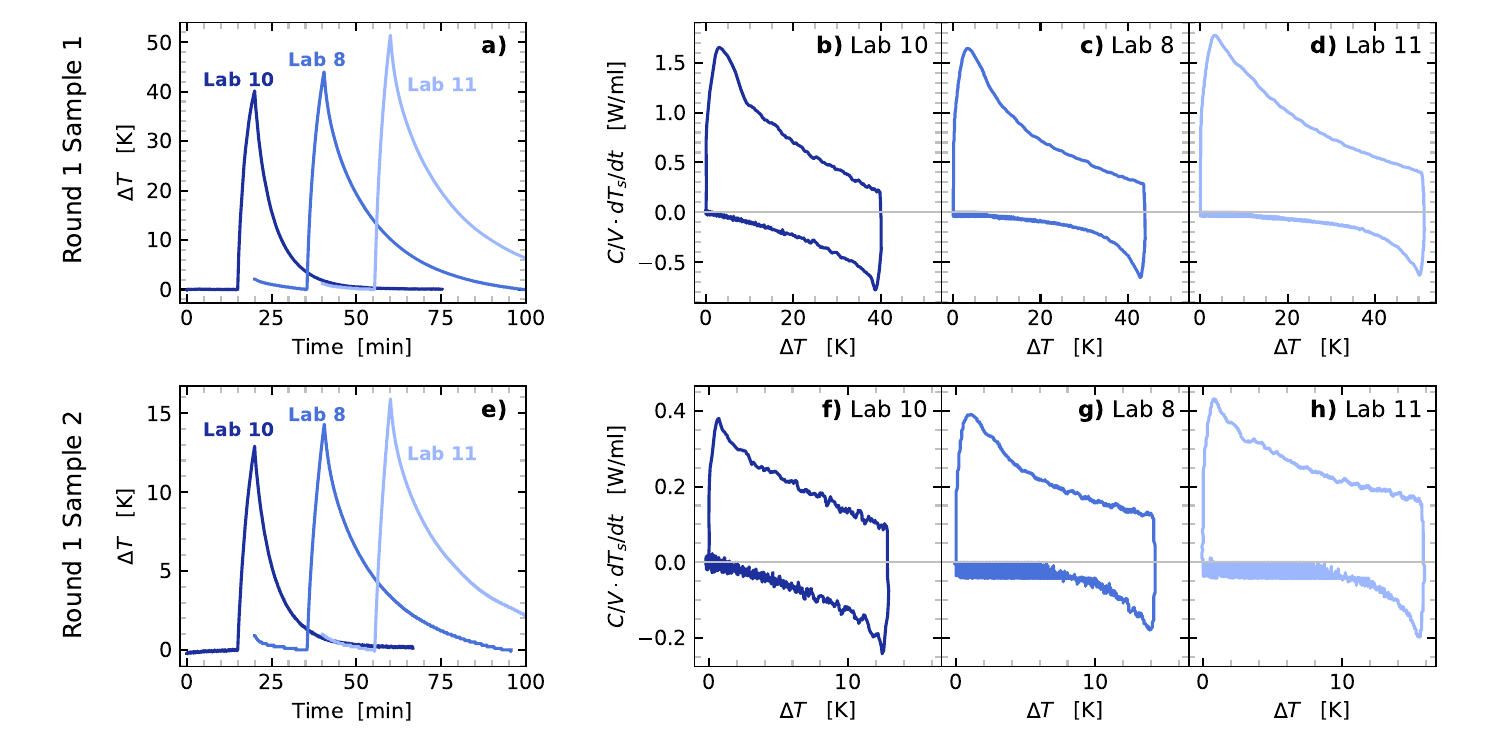}
    \caption{Examples of data from three laboratories in which apparent non-linear heat losses are observed. The data are from round 1 and measured on Samples 1 (top panels) and 2 (bottom panels). \textbf{a)} and \textbf{e)} Show measured temperature change as a function of time. \textbf{b, c, d, f, g,} and \textbf{h)} Show the corresponding slope curves.}
    \label{fig:Non linear heat loss}
\end{figure*}

\begin{table*}[b!]
    \centering
    \begin{tabular}{|c|c|c|c|c|c|c|c|c|c|c|c|c|c|c|c|c|c|c|c|c|c|c|}
    \hline
        \rowcolor{lightgray!20}
        Laboratory & 1 & 2 & 3 & 4 & 5 & 6 & 7 & 8 & 9& 10& 11& 12& 13& 14& 15& 16& 17& 18& 19& 20& 21 \\
        \hline
        Temp. resolution S1 & & \cellcolor{violet!50} & & & & & & & & & & & & & & & & & & & \\
        \arrayrulecolor{gray!20}\specialrule{.01em}{0em}{0em} 
        \arrayrulecolor{black}
        Temp. resolution S2 & \cellcolor{violet!50} & \cellcolor{violet!50} & & & &  & & & & & & & & \cellcolor{violet!50} & \cellcolor{violet!50}& & \cellcolor{violet!50}& & & & \\
        \arrayrulecolor{gray!20}\specialrule{.01em}{0em}{0em} 
        \arrayrulecolor{black}
        Field sensitive therm. & \cellcolor{RedViolet!80} & & & & & & & & & & & & & & & \cellcolor{RedViolet!80} &  & & & \cellcolor{RedViolet!80} &  \\
        \arrayrulecolor{gray!20}\specialrule{.01em}{0em}{0em}
        \arrayrulecolor{black}
         Heat oscillations S1 & & & & & & \cellcolor{purple!65} & & & & & &  & & & & & & & & &  \\
        \arrayrulecolor{gray!20}\specialrule{.01em}{0em}{0em}
        \arrayrulecolor{black}
        Heat oscillations S2 & & & & & \cellcolor{purple!65} & \cellcolor{purple!65} & \cellcolor{purple!65} & & & & & & \cellcolor{purple!65} & & & & & & & &  \\
        \arrayrulecolor{gray!20}\specialrule{.01em}{0em}{0em} 
        \arrayrulecolor{black}
        Other exclusions & & & & & & & & & & & & & & & & & & & & \cellcolor{blue!50} & \cellcolor{blue!50}\\
        \hline
       S1 & \cellcolor{gray!50} & \cellcolor{gray!50} & & & & \cellcolor{gray!50}& & & & & & & & & & \cellcolor{gray!50} & & & & \cellcolor{gray!50} &\cellcolor{gray!50}\\
       \arrayrulecolor{gray!20}\specialrule{.01em}{0em}{0em} 
        \arrayrulecolor{black}
        S2 & \cellcolor{gray!50} & \cellcolor{gray!50} & & & \cellcolor{gray!50} & \cellcolor{gray!50}&\cellcolor{gray!50} & & & & & &\cellcolor{gray!50} & \cellcolor{gray!50}& \cellcolor{gray!50} & \cellcolor{gray!50} & \cellcolor{gray!50} & & & \cellcolor{gray!50} &\cellcolor{gray!50}\\
        \hline
       
    \end{tabular}
    \caption{Summary of all measurement rejections. Coloured boxes show rejections at specific requirements, and the grey boxes in the last two rows show an overall rejection summary for Sample 1 (S1) and Sample 2 (S2). }
    \label{tab:rejection summary}
\end{table*}

A linear heat loss is required for using the corrected slope method to determine the heating power, $P_\text{MNP}$, since $P_\text{loss}=L\Delta T$ is a crucial assumption behind this method. It is therefore important to investigate the prevalence of apparent non-linear heat losses, especially since the corrected slope method was used to determine ILP in the RADIOMAG study \autocite{wells2021challenges}.

\cref{fig:Non linear heat loss} shows examples of measurements from laboratories where apparent non-linear heat losses are observed. If the heat loss were linear, the slope curves would depend linearly on $\Delta T$ as illustrated in \cref{fig:slopeCurve}. Instead, the slope curves in \cref{fig:Non linear heat loss} are clearly non-linear in both the heating and cooling phases. Comparing \cref{fig:Non linear heat loss} b-d vs. f-h shows that the non-linearity is prevalent for both Sample 1 and Sample 2. Thus, the non-linearity does not appear dependent on the maximal temperature reached in the sample.

It has recently been shown that slope curves with similar non-linear shapes as the ones observed in \cref{fig:Non linear heat loss} can result from an increase in the temperature of the sample insulation \autocite{Hanson2023impact}. 
To this end, the non-linear shape of the slope curves could be fully explained by the heat loss being larger in the heating phase than in the cooling phase at any given value of $\Delta T$ due to the increasing temperature in the sample insulation over time. 

The non-linearity observed in \cref{fig:Non linear heat loss} can presumably be explained by the use of sample insulation, as laboratories 8 and 11 used Dewars with a vacuum for the sample holder and insulation, while laboratory 10 used a 0.5 cm thick nylon cylinder. Generally, sample insulation is found to be widespread in the RADIOMAG study, with at least 17 of the 20 non-adiabatic laboratories using some sort of insulation. This is a conservative number, since some laboratories only provided limited information on their sample insulation. 

Out of the 20 non-adiabatic setups, 7 (laboratories 7, 8, 9, 10, 11, 17, and 18) are observed to lead to apparent non-linear heat losses, all of which used sample insulation. Only laboratory 15 was observed to exhibit a linear heat loss, although they also used insulation. 
For the remaining 12 non-adiabatic setups, the linearity of the heat loss could not be determined due to other issues, such as field-sensitive thermometers, poor temperature resolution, or non-physical heat oscillations.


Based on the strong prevalence of non-linear heat losses across laboratories, we recommend that when using the corrected slope method, it should be standard practice to verify the linearity of the heat loss through slope curves. If the slope curves and the heat loss are found to appear non-linear, we suggest this might be caused by heating in the sample environment. Thus, to mitigate the apparent non-linear heat loss, the sample insulation can be removed, and a constant external temperature can be maintained by passing a controlled air flow across the sample. Alternatively, the temperature inside the insulation near the sample can be measured and used to correctly define $\Delta T$ \autocite{Hanson2023impact}.

\begin{figure*}[t]
    \centering
    \begin{subfigure}[b]{0.45\textwidth}
        \centering
        \includegraphics[width=\textwidth]{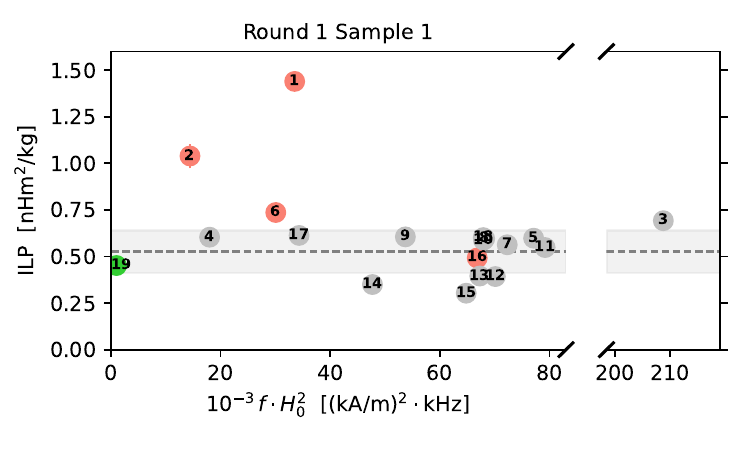}
    \end{subfigure}%
    ~ 
    \begin{subfigure}[b]{0.45\textwidth}
        \centering
        \includegraphics[width=\textwidth]{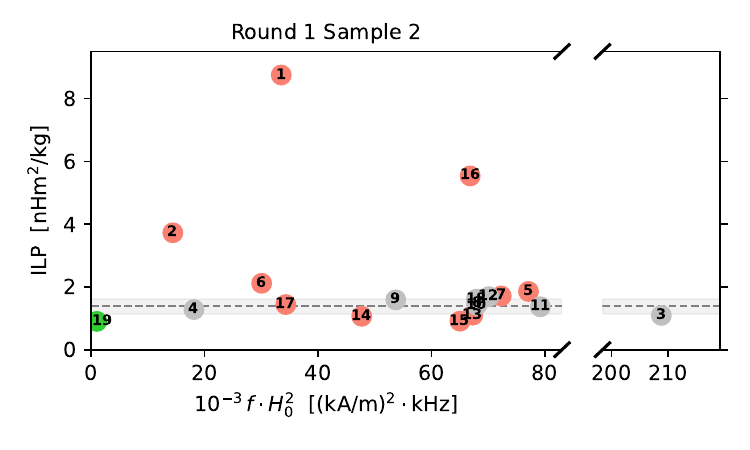}
    \end{subfigure}
    ~ 
    \begin{subfigure}[b]{0.45\textwidth}
        \centering
        \includegraphics[width=\textwidth]{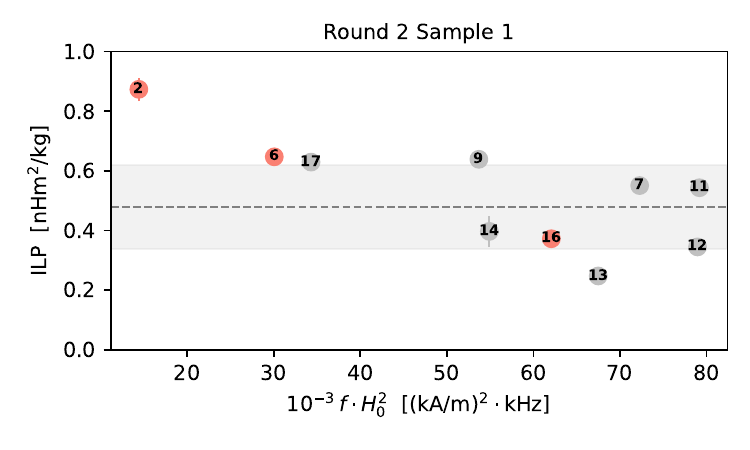}
    \end{subfigure}
    ~ 
    \begin{subfigure}[b]{0.45\textwidth}
        \centering
        \includegraphics[width=\textwidth]{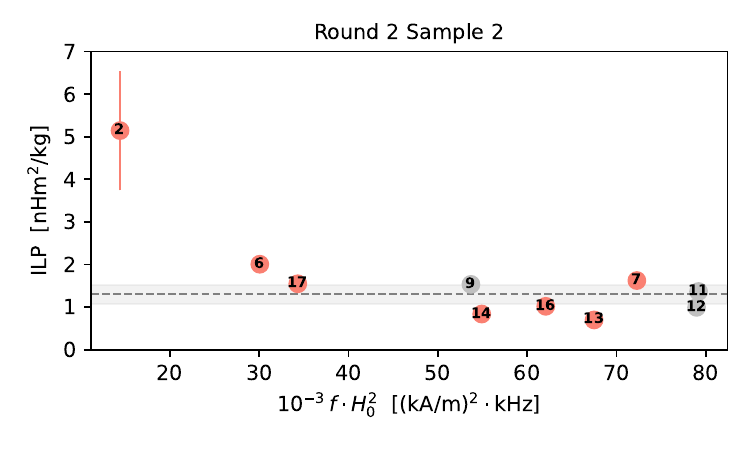}
    \end{subfigure}
    
    \caption{Distribution of ILP values estimated with the initial slope method.
    \sffamily \textcolor{lightgray}{\textbf{Grey}}\rmfamily: Accepted non-adiabatic laboratories, 
    \sffamily \textcolor{Green}{\textbf{green}}\rmfamily: Adiabatic laboratory, and 
    \sffamily \textcolor{red!70}{\textbf{red}}\rmfamily: Rejected laboratories. The grey dashed line indicates the mean ILP value of all the accepted laboratories and the grey filled area indicates their standard deviation.}
    \label{fig:ILP vs H2f}
\end{figure*}

\subsection{Re-estimation of ILP values}
Here, ILP values of the RADIOMAG data are re-estimated from slope curves using the initial slope method (cf. \cref{fig:slopeCurve}), which is not affected by the (non-)linearity of the heat loss. It is of interest to re-estimate the ILP values because the previous estimates were based on the corrected slope method \autocite{wells2021challenges}, which is problematic for at least 7 of the 20 non-adiabatic setups due to non-linear heat losses. Thus, it is interesting see how the change in analysis method affects the deviations in estimated ILP-values. Additionally, we investigate how the exclusion criteria (temperature resolution, field-sensitive thermometers, and non-physical heat oscillations) affect the distribution of the ILP values.


\subsubsection{Exclusions of laboratories}
\begin{table*}[b]
    \centering
    \begin{tabular}{ccccccc}
    \hline
    \rowcolor{lightgray!20}
    & &\Gape[0pt][2pt]{\makecell{Number of \\ accepted labs}} 
    &  \Gape[0pt][2pt]{\makecell{$\sigma_a$ Initial slope \\ *no rejections }}
    & \makecell{$\sigma_b$ Initial slope \\ with rejections }  
    & \makecell{$\sigma_c$ Corrected slope \\ Wells et al. \autocite{wells2021challenges} } 
    & \\ \hline
    & Round 1 Sample 1 & 15 out of 21 & 30 \% & 22 \%  & 31 \% & \\
    & Round 1 Sample 2 & 9 out of 21 & 63 \% (23 \% lab2+16) & 18 \%  & 28 \%  &\\ 
    \arrayrulecolor{gray}\hline
    & Round 2 Sample 1 & 7 out of 12 & 34 \% & 30 \%  & 32 \% & \\ 
    &Round 2 Sample 2 & 3 out of 12 & 77 \% (32 \% lab2) & 24 \%  & 39 \% &\\ 
    \arrayrulecolor{black}\hline
    \end{tabular}
    \caption{ILP standard deviation overview across the different rounds and samples. *No rejections means all labs included in Wells et al. \autocite{wells2021challenges} are included in the estimate of $\sigma_a$. The values in parentheses state the standard deviation if only laboratory 2 and 16 are rejected. An overview of the rejections is seen in \cref{tab:rejection summary}.}
    \label{tab:deviations}
\end{table*}

\cref{tab:rejection summary} provides an overview of the laboratories excluded from the ILP re-estimation. 
In addition to the exclusion criteria described in the previous sections, the data from laboratory 20 is excluded since the sample was diluted, leading to sample damage. Laboratory 21 is excluded because they reported a field amplitude of 14-17 mT corresponding to a 19\% deviation in the field amplitude and a 39 \% deviation when calculating ILP. 
Data from laboratory 20 and 21 are completely excluded from the rest of the analysis. 

\subsubsection{ILP values by initial slope}
\cref{fig:ILP vs H2f} shows all re-estimated ILP values as a function of $fH_0^2$. The grey data points represent the accepted non-adiabatic measurements, the green data point is the adiabatic measurement, and the red data points are the rejected measurements. The grey dashed lines show the mean ILP value of all the accepted laboratories, and the grey filled areas indicate the corresponding standard deviation.

It is observed that the accepted (grey) points remain rather constant as a function of $fH_0^2$ in all four plots in \cref{fig:ILP vs H2f}, confirming the validity of linear response theory for both samples \autocite{carrey2011simple, wildeboer2014reliable}. This confirms that ILP is a reliable metric for inter-laboratory comparison for both samples.

The excluded (red) data points include strong outliers (labs 1, 2, and 16) but also a few points (labs 7 and 17) that align well with the mean ILP value. Overall, the quality criteria efficiently remove data points that deviate significantly from the mean ILP value.

\cref{tab:deviations} provides an overview of the ILP standard deviations, $\sigma$, for following three different cases 
\begin{enumerate}[label={$\sigma_\alph*$)}]
    \item  Initial slope re-estimation with no rejections, i.e., $\sigma_a$ is calculated based on the same laboratories as accepted in Wells et al. \autocite{wells2021challenges} where only laboratory 1 is rejected.
    \item   Initial slope re-estimations where laboratories are rejected according to \cref{tab:rejection summary}.
    \item  A reprint of the standard deviations determined in Wells et al. based on the corrected slope method \autocite{wells2021challenges}.
\end{enumerate}

\begin{figure*}[b!]
    \centering
    \includegraphics[width=\linewidth]{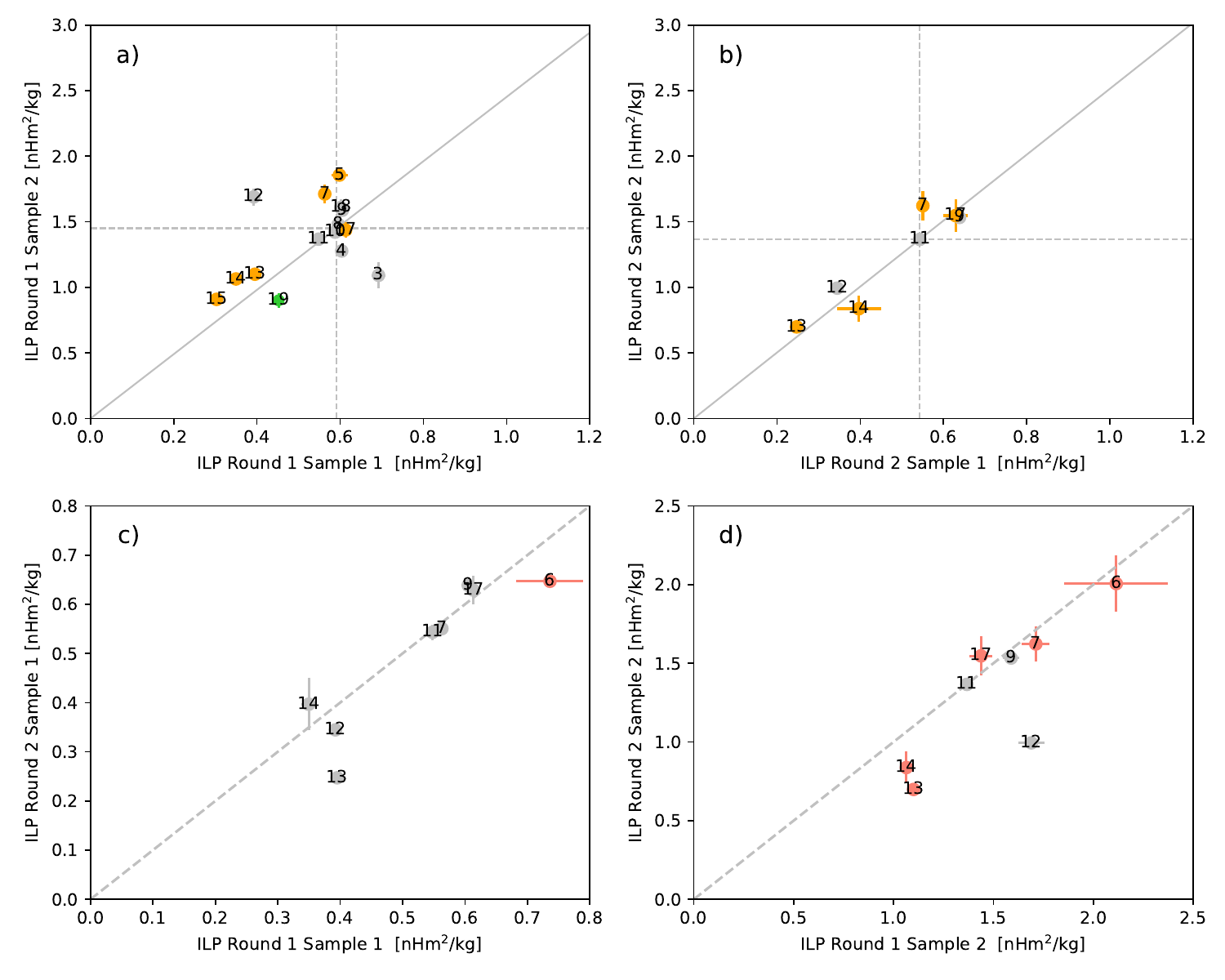}
    \caption{Comparison of ILP values from different rounds and samples. \textbf{a) and b)} are Youden plots for Samples 1 and 2, respectively. These plots are used to investigate whether systematic or random errors are dominant.
    \sffamily\textcolor{lightgray}{\textbf{Grey}}\rmfamily: Measurement accepted for both Samples 1 and 2.
    \sffamily \textcolor{orange}{\textbf{Orange}}\rmfamily: Measurement only accepted for Sample 1 and not for Sample 2.
    \sffamily\textcolor{Green}{\textbf{Green}}\rmfamily: Adiabatic measurements.
    The vertical and horizontal dashed lines indicate the median ILP values of the accepted data points. The diagonal solid grey line has a slope such that it intersects the Manhattan median. 
    \textbf{c) and d)} compare the ILP values of the same sample across the rounds to investigate whether the refinement of the protocol and the repositioning of the samples changed the ILP values.
    \sffamily\textcolor{lightgray}{\textbf{Gray}}\rmfamily: Measurements accepted for the given sample in both rounds. 
    \sffamily \textcolor{red!70}{\textbf{Red}}\rmfamily: Measurements rejected in both rounds. The diagonal dashed line has a slope of 1, representing the line where the ILP values have not changed between the two measurement rounds. 
    Measurements from laboratory 1, 2, and 16 are excluded due to their strong deviations caused by the field-sensitive thermometers or insufficient temperature resolution. 
    }
    \label{fig:ILP vs ILP}
\end{figure*}

\cref{tab:deviations} shows that the re-estimated standard deviations after applying the rejection criteria are consistently lower than those reported by Wells et al. \autocite{wells2021challenges}. The post-exclusion standard deviations, $\sigma_b$, range from 18\% to 30\%, representing a reduction of up to 38\% compared to the original values reported in Wells et al. \autocite{wells2021challenges}. Thus, the combination of data rejection and the use of the initial slope method significantly decreases inter-laboratory variability. However, even these reduced deviations remain problematically high, suggesting that while the exclusion criteria are effective, additional sources of variability remain unaccounted for.

In both rounds, the largest relative reduction in standard deviation, when comparing $\sigma_c$ with $\sigma_b$, is observed for Sample 2 (28\% $\rightarrow$ 18\% and 39\% $\rightarrow$ 24\%). 
This is caused by Sample 2 having a lower heating power, making it more sensitive to the identified instrumentation issues of poor temperature resolution, field sensitivity, and non-physical heat oscillations.

Finally, \cref{tab:deviations} shows that the standard deviation without rejections, $\sigma_a$, can be as high as 77\%, primarily due to the inclusion of data from laboratories 2 and 16. When only these two laboratories are excluded, the standard deviation drops significantly (e.g., to 32\%), as indicated in parentheses in the table. 




\subsubsection{Systematic or random errors}

In order to further explore the variations and possible inconsistencies in the RADIOMAG dataset, we have used the so-called Youden plots. In essence, this type of graphical representation compares the inter-laboratory differences with the intra-laboratory ones and is used for investigating whether the errors across the laboratories are random or systematic \autocite{Youden1972}.

\cref{fig:ILP vs ILP}a and b show Youden plots for rounds 1 and 2, respectively. The scatter points show the ILP values of Sample 1 vs. Sample 2 for a given laboratory. The dashed grey lines show the medians of the accepted ILP values for each sample. In the ideal case with no error, all points would lie at the intersection of these lines, known as the Manhattan median.
The solid grey line is called the 45$\degree$ line, which passes through the origin and the Manhattan median. A distribution of points along this line suggests systematic errors, while a circular distribution around the Manhattan median suggests random errors. 
Grey points represent data accepted in rounds 1 and 2, while the orange points are accepted only in round 1. Therefore, the y-values of the orange points should be interpreted with caution.

\cref{fig:ILP vs ILP}a shows the Youden plot for round 1. 6 out of 8 fully accepted (grey) data points lie close to the Manhattan median. 
Only two of the grey points (labs 3, 12) deviate strongly, and their deviation is perpendicular to the 45$\degree$ line. This suggests either random error or sample-specific measurement issues (e.g., underestimating Sample 1 and overestimating Sample 2 due to differences in heating power)
The green point (lab 19), representing the adiabatic setup, also deviates from the Manhattan median but lies along the 45$\degree$ line, indicating a potential systematic error. 

The orange points represent laboratories accepted only for Sample 1, and thus, their Sample 2 ILP values may contain large errors. 
The data points for laboratories 5, 7, and 17 are already close, or can be brought close to the Manhattan median by shifting the Sample 2 ILP value. 
However, for laboratories 13, 14, and 15, the points cannot be brought close to the Manhattan median regardless of the Sample 2 value. 
This suggests that these laboratories contribute significantly to the ILP deviation of Sample 1, and not just Sample 2. The criteria used in this study were not able to identify the cause of these deviations.

\cref{fig:ILP vs ILP}b shows the Youden plot for round 2, which includes fewer data points due to lower participation (12 laboratories) in this round.
With limited data, statistical conclusions are less robust. 
However, the median ILP values remain similar to those in round 1 (within 8\%), and most laboratories occupy similar positions in both rounds. Only laboratory 12 shows a significant shift. 
This could be due to (i) changes in measurement procedure following protocol refinement in round 2, or (ii) repositioning of the sample in a setup that is sensitive to placement errors. Notably, samples were not repositioned within individual rounds, but were repositioned between rounds.

\cref{fig:ILP vs ILP}c and d compare ILP values for the same sample across different rounds. These plots assess whether ILP values are reproducible between rounds or affected by protocol changes or sample repositioning. The dashed grey line indicates perfect agreement between rounds. 
Most data points, both fully accepted (grey) and fully rejected (red), lie close to this line, indicating good in-house reproducibility. 
This suggests that protocol refinements and sample repositioning generally did not affect ILP measurements.
The few points that deviate more strongly (labs 6, 12, and 13) tend to lie below the dashed line, indicating slightly lower heating power measured in round 2. 
No specific cause for this trend has been identified.

In conclusion, most accepted data points show good agreement across samples and measurement rounds. A small number of laboratories (3, 12, 13, 14, 15, and 19) appear to be the primary contributors to the observed deviations. However, the underlying causes of their deviations could not be identified in this study.

\subsection{Protocol}
\label{sec:protocol}
In light of the analysis made,  we propose the following five-step diagnostic protocol to ensure reliable and reproducible ILP measurements using AC calorimetry. The protocol aims to help identify experimental challenges and improve the AC calorimetric setup proactively before attempting to measure the heating power of a sample. 
To facilitate the use of the protocol, we have developed a Jupyter Notebook for plotting and investigating slope curves. The notebook is available at GitHub alongside a guide for installation and how to use the code for research without previous programming experience: \href{https://github.com/LiseHanson/Slope-Curves---Jupyter-Notebook}{link} \autocite{GitHubSlopeCurves}. 

\textbf{Step 1)} Avoid premature rounding errors by ensuring that the thermometer records more digits than the resolution stated for the temperature sensor. This prevents artificial discretization of the temperature signal and improves the accuracy of numerical derivatives. 

\textbf{Step 2)} Characterize the thermometer for field sensitivity and noise level by measuring the temperature response of a non-heating sample (e.g., water or another non-magnetic, non-conductive material with similar thermal mass as the target sample). From the resulting data, generate a slope curve.  
First, inspect the shape of both the raw data and the slope curve to identify any AC field-induced artefacts, as illustrated in \cref{fig:Thermometer sensitivity to AC field}. Then, quantify the maximum value of $C \cdot dT_s/dt$ to estimate the noise level (for field-insensitive thermometers) or the systematic error (for field-sensitive thermometers), expressed in units of W.  

\textbf{Step 3)} Perform AC calorimetric measurements on a sample with magnetic nanoparticles capable of generating heat by induction. Investigate the resulting slope curves for non-physical temperature oscillations and non-linear heat loss. These artefacts can significantly affect the accuracy of heating power estimation and must be identified and solved prior to analysis. 

\begin{figure}[t]
    \centering
    \includegraphics[width=\linewidth]{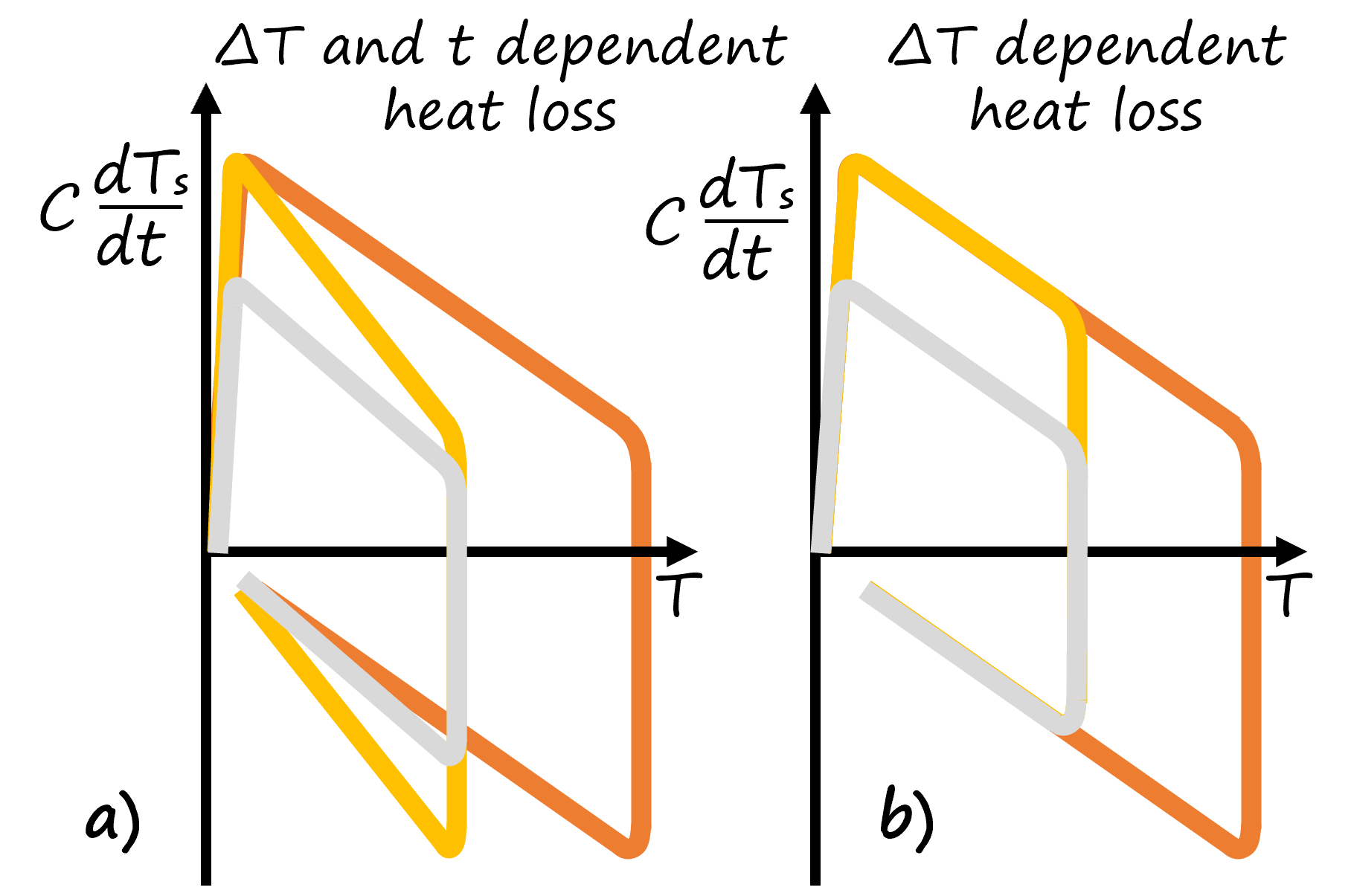}
    \caption{Illustration of how slope curves with different field parameters (field application time, field amplitude, and frequency) can be used to compare the heat loss in the cooling phase. \textbf{a)} Slope curves where the heat loss is not purely dependent on temperature but also dependent on time. \textbf{b)} Slope curves where the heat loss only depends on temperature.}
    \label{fig:protocol different fields test time dependence}
\end{figure}

\textbf{Step 4)} Perform AC calorimetric measurements on the same sample at different field settings (field application time, field amplitude, and/or frequency) while maintaining all other parameters the same (sample volume, cooling flow in the coil, room temperature, etc.). Compare the slope curves of these different measurements. If the heat loss is purely temperature dependent, the cooling phase should overlap for all measurements, see \cref{fig:protocol different fields test time dependence}. For these comparison measurements, it is important that the sample is thermally equilibrated before the measurement is initialised, and the room temperature is the same. 

If the heat loss is not purely temperature dependent and linear, an attempt should be made to change this, such that the corrected slope method can be applied.
Based on our previous studies, we suspect the apparent non-linear heat losses appear due to heating of the sample's immediate surroundings, typically the insulation \autocite{Hanson2023impact}. 
To mitigate these apparent non-linear heat losses, we suggest two approaches: 1)  Maintaining a constant temperature around the sample by removing the insulation and passing an air flow across the sample. 2) Correctly defining $\Delta T$ as the difference between the sample temperature and the insulation temperature. If the sample environment temperature is not maintained at a known fixed temperature, this requires measuring the temperature at both positions (sample and insulation)\autocite{Hanson2023impact}.

\textbf{Step 5)} Estimate the heating power of the sample from slope curves using an appropriate method, i.e., the corrected slope method or the initial slope method. The corrected slope method is preferred over the initial slope method, assuming a linear heat loss could be implemented and confirmed in step 4. This preference is based on an expected higher validity of the corrected slope method \autocite{wildeboer2014reliable,wells2021challenges}. Thus, the initial slope method should only be used if non-linear heat loss cannot be remedied. 

Compare the estimated heating power to the noise and systematic error levels identified in Step 2 to assess the reliability of the measurement. For the corrected slope method, follow the protocol suggested by Wildeboer et al. \autocite{wildeboer2014reliable}. Additionally, the temperature homogeneity in the sample can be confirmed by measuring temperatures at multiple positions in the sample in experiments of the same style as in \autocite{wildeboer2014reliable} and \autocite{Hanson2023impact}.
\section{Conclusions}
In this study, four prevalent instrumentation issues in AC calorimetry were identified: insufficient temperature resolution (5 of 21 laboratories), AC field-sensitive thermometers (3 of 21 laboratories), non-physical temperature oscillations (4 of 21 laboratories), and apparent non-linear heat loss ($\geq$7 of 20 laboratories). 
Based on these findings, data from laboratories exhibiting these issues were excluded, and ILP values were re-estimated using only measurements of sufficient quality.

The initial slope method was employed for ILP re-estimation, as the corrected slope method relies on the assumption of linear heat loss, which could not be confirmed in the majority of the laboratories. 
However, it must be underlined that we encourage and prefer the use of the corrected slope method over the initial slope method if a linear heat loss is validated in future measurements. We expect that such linear losses can be implemented either by mitigating temperature increments in the near sample surroundings by removing the sample insulation and passing an air flow across the sample, or by additionally measuring the temperature in the sample insulation to correctly define the temperature difference, $\Delta T$.

Applying the rejection criteria, the re-estimated ILP values show a deviation of 18–30\%, representing a reduction of up to 38\% compared to the deviation reported by Wells et al. 
The criteria were found to effectively exclude ILP values that deviated significantly from the median of the accepted dataset. 
Moreover, high reproducibility between measurement rounds was observed, indicating that sample repositioning within instruments generally does not introduce random errors.

Despite the improvement, the remaining deviation of 18–30\% remains substantial and warrants further investigation. 
Youden plot analysis revealed that six laboratories (labs 3, 12, 13, 14, 15, and 19) were the primary contributors to the residual deviation, exhibiting a mix of random and systematic errors. 
The underlying causes of these deviations were not identified in this study and should be explored in future work.

A promising future direction would be to repeat a round-robin study similar to the RADIOMAG study, but with laboratories instructed to ensure linear heat losses in their measurements, allowing the corrected slope method to be applied when comparing ILP values. We expect that proper use of the corrected slope method could further reduce the ILP deviations, compared to using the less robust initial slope method, which was required in this study due to apparent non-linear heat losses.

To support improved data quality in future AC calorimetry studies, a diagnostic protocol is proposed in \cref{sec:protocol}. This protocol is based on the use of slope curves, which provide a powerful visual tool for identifying measurement artefacts. We recommend that laboratories adopt this protocol to assess whether any of the identified instrumentation issues are present in their setups and do this proactively to establish reliable measurements in the acquisition phase.

\section{Author contributions}
Lise G. Hanson: Conceptualisation, formal analysis, investigation, methodology, software, visualisation, writing – original draft. 
Daniel Ortega: Methodology, resources, writing – review. 
Catrine Frandsen: Conceptualisation, funding acquisition, supervision, writing – review \& editing.

\section{Conflicts of interest}
There are no conflicts to declare.

\section{Data availability}
This study was carried out using publicly available data from the RADIOMAG Inter-laboratory Comparison of Magnetic Hyperthermia Characterisation Measurements. These data are available at \url{https://zenodo.org/records/4281154} with DOI: \href{https://doi.org/10.5281/zenodo.4281153}{10.5281/zenodo.4281153}.

\section{Software availability statement}
A Jupyter Notebook intended for easy implementation of slope curve analysis is freely available at \url{https://github.com/LiseHanson/Slope-Curves---Jupyter-Notebook} \autocite{GitHubSlopeCurves}. The notebook is aimed at scientists without prior experience with programming. A guide on how to use the notebook is provided inside the notebook. At the GitHub page, a guide on installing Python and JupyterLab is provided. It is expected that installation of the software and learning to use the script takes only 1 to 2 hours. 

\section{Acknowledgement}
We thank the RADIOMAG consortium for access to their data. This work has been supported by the National Committee for Research Infrastructure (NUFI) through the ESS Lighthouse programme on Quantum materials and the Independent Research Fund Denmark grant 4283-00355B. Additionally, we thank James Wells, Quentin Pankhurst and Paul Southern for fruitful discussions. 

\printbibliography[heading=subbibliography]

\end{refsection}

\clearpage
\setcounter{page}{1}
\setcounter{figure}{0}
\setcounter{table}{0}
\setcounter{equation}{0}
\setcounter{section}{0}
\renewcommand{\thefigure}{SI~\arabic{figure}}
\renewcommand{\thetable}{SI~\arabic{table}}
\renewcommand{\theequation}{SI~\arabic{equation}}
\renewcommand{\thesection}{SI~\arabic{section}}

\onecolumn 
\begin{refsection}
\begin{center}
\Large{\textbf{Unravelling Challenges in Heating Power Measurements for Magnetic Hyperthermia - the RADIOMAG Round Robin Study Revisited}}\\
\normalsize
\vspace{3mm}
\textbf{\today}\\
\vspace{3mm}
Lise G. Hanson$^a$\orcidlink{0000-0002-8710-2610}, 
Daniel Ortega$^{b,c}$ \orcidlink{0000-0002-7441-8640},
Cathrine Frandsen$^{a,*}$\orcidlink{0000-0001-5006-924X}\\
\vspace{3mm}
$^a$Department of Physics, Technical University of Denmark, 2800 Kgs. Lyngby, Denmark\\
$^b$Condensed Matter Physics Department, University of Cádiz, 11510 Puerto Real, Spain\\ 
$^c$Biomedical Research and Innovation Institute of Cádiz, University of Cádiz, 11009 Cádiz, Spain\\
$^*$Corresponding authors, e-mail: \href{fraca@fysik.dtu.dk}{fraca@fysik.dtu.dk}
\end{center}

\section{Slope curves for all laboratories}
\label{SI:Slope curves}
The slope curves are obtained with numerical differentiation through a Savitzky–Golay filter with a polynomial order of 1 and a window width of 15 s. 
\foreach \n in {1,...,21}{
\begin{figure}[H]
    \centering
    \includegraphics[width=0.7\linewidth]{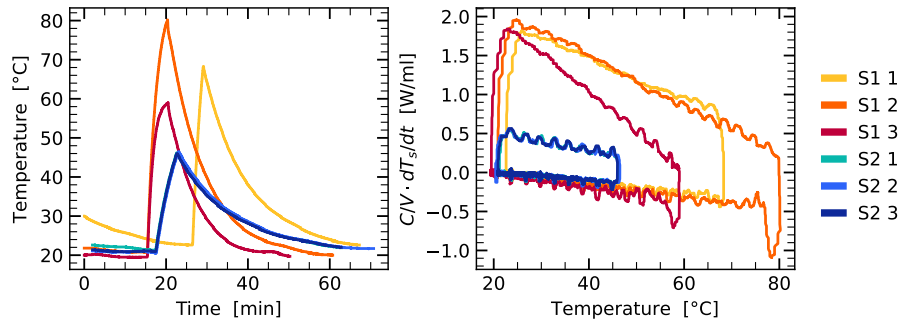}
    \caption{AC calorimetry data for laboratory \n $\,$ measured on sample 1 (S1) and sample 2 (S2). Both samples have been measured repeatedly three times. 
    Left: Measured sample temperature as a function of time. Right: Slope curve calculated from the data on the left. }
    \label{SIfig:Slope curve lab \n}
\end{figure}
}

\section{Temperature resolution criteria}
\label{SI:T res criteria}
Table or plot of all $\delta T_s$ values and limit values. 
\begin{figure}[H]
    \centering
    \includegraphics[width=\linewidth]{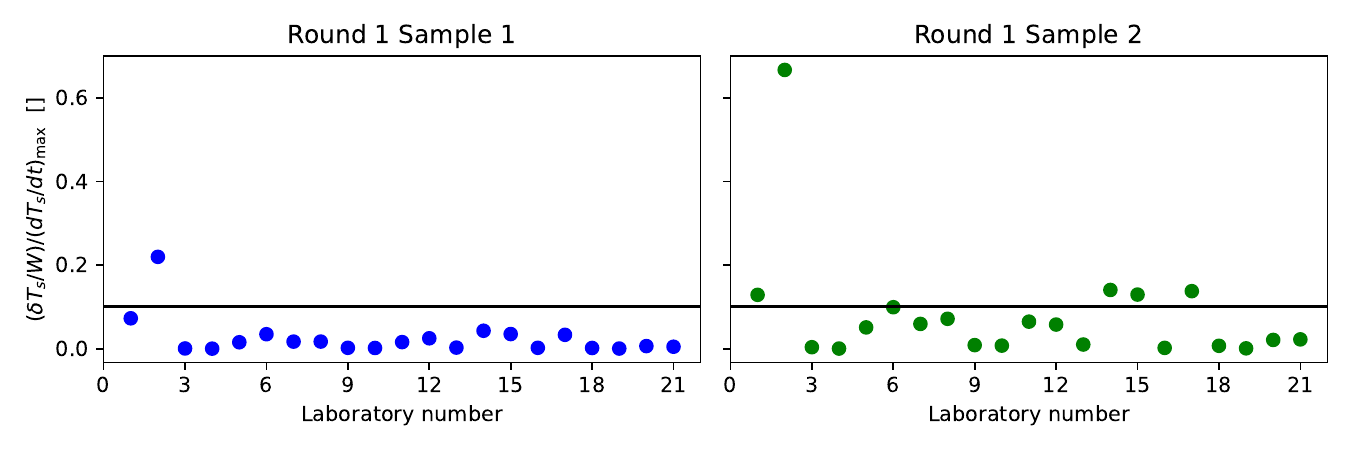}
    \caption{Overview of temperature criteria for all laboratories for round 1 sample 1 (\textbf{a}) and 2 (\textbf{b}). The criteria $ \left. \frac{\delta T_s}{W}\middle/\left[\frac{dT_s}{dt}\right]_\text{max} \right. <0.1$ is shown with the black line.}
    \label{fig:Tres_limit}
\end{figure}

\section{ILP estimates adiabatic measurements}
\label{SI:ILP adiabatic}

\begin{figure}[H]
    \centering
    \includegraphics[width=0.7\linewidth]{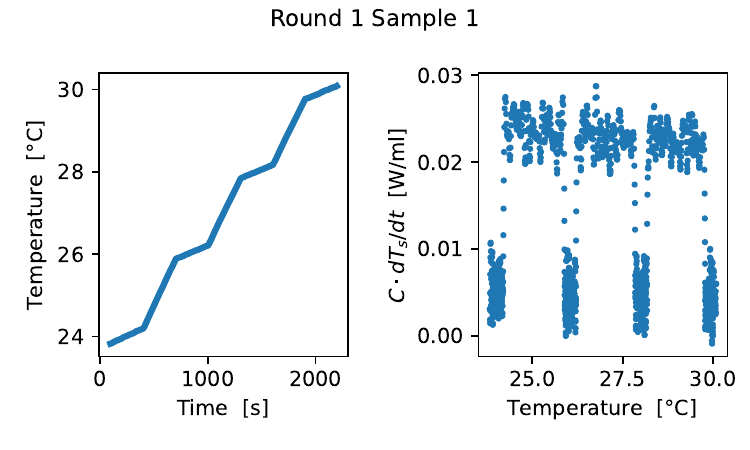}
    \caption{ILP determination of the adiabatic laboratory 19. Left show the raw data with sample temperature as function of time. The sample is exposed to three consecutive field pulses. Right: slope curve obtained from the data on the left. The three high plateaus are the field on phases and the four lower plateau are the field off phases. }
    \label{fig:ILP 19 method}
\end{figure}
The heating power of the adiabatic laboratory is determined by 
\begin{equation}
    P_\text{MNP}/V = \frac{C}{V}\left( \left\langle\frac{dT_s}{dt}_\text{Field on}\right\rangle -  \left\langle\frac{dT_s}{dt}_\text{field off} \right\rangle \right)
    \label{eq:adiabatic ILP}
\end{equation}
Where $\langle\rangle$ indicates the average across one plateau and $V$ is the sample volume. 
Three estimates of ILP are obtained from \cref{eq:adiabatic ILP} by using one field on plateau and the subsequent field off plateau. 
\begin{gather}
    \text{ILP} = \frac{P_\text{MNP}}{m_\text{MNP}}\frac{1}{H^2 f} = \frac{P_\text{MNP}}{V}\frac{1}{c}\frac{1}{H^2 f}.
\end{gather}

\printbibliography[heading=subbibliography]
\end{refsection}

\end{document}